\documentclass[letterpaper,12pt]{article}

\usepackage{fancyheadings}
\usepackage[dvips]{graphicx}
\usepackage{graphpap}
\usepackage{ifthen}
\usepackage{enumerate}
\usepackage{cite}
\usepackage{amssymb}
\usepackage[mathscr]{eucal}
\usepackage[errorshow]{tracefnt}
\usepackage{fancybox}
\usepackage{amsfonts}
\usepackage{amsmath}
\usepackage{amssymb}
\usepackage{amsthm}
\usepackage{eucal}
\usepackage{eufrak}

\newcommand{\cM}{\ensuremath{\mathcal{M}}}
\newcommand{\cN}{\ensuremath{\mathcal{N}}}
\newcommand{\cP}{\ensuremath{\mathcal{P}}}
\newcommand{\cQ}{\ensuremath{\mathcal{Q}}}

\newcommand\nn{\nonumber}

\newcommand{\bor}{U}

\newcommand{\beq}{\begin{equation}}
\newcommand{\beqn}{\begin{equation*}}
\newcommand{\eeq}{\end{equation}}
\newcommand{\eeqn}{\end{equation*}}
\newcommand{\beqa}{\begin{eqnarray}}
\newcommand{\beqan}{\begin{eqnarray*}}
\newcommand{\eeqa}{\end{eqnarray}}
\newcommand{\eeqan}{\end{eqnarray*}}
\newcommand{\bdm}{\begin{displaymath}}
\newcommand{\edm}{\end{displaymath}}

\newcommand{\la}{\langle}
\newcommand{\ra}{\rangle}

\newtheorem{theorem}{Theorem} 
\newtheorem{lemma}[theorem]{Lemma}

\newtheorem{cor}[theorem]{Corollary}

\def\Pf{\noindent \textbf{Proof. }}

\def\der'{\mathfrak{der}'\,}
\def\der{\mathfrak{der}\,}
\def\str'{\mathfrak{str}'\,}
\def\str{\mathfrak{str}\,}

\def\g{\gamma}

\def\g{\mathfrak{g}}
\def\so{\mathfrak{so}}

\def\sl{\mathfrak{sl}}

\def\qed{\hspace{\stretch{1}} $\square$ \\
\noindent}

\newcommand{\al}{\alpha}
\newcommand{\be}{\beta}

\newcommand{\de}{\delta}
\newcommand{\ga}{\gamma}

\newcommand{\dlb}{\ensuremath{[\![}}
\newcommand{\drb}{\ensuremath{]\!]}}

\numberwithin{equation}{section}

\begin{document}

\hfill{\tt ULB-TH/11-22}\\
\vskip-10pt
\hfill {\tt \today}

\pagestyle{empty}

\vspace*{3.5cm}

\noindent
\begin{center}
{\LARGE {\sf \textbf{Tensor hierarchies, Borcherds algebras and $E_{11}$}}}\\
\vspace{.3cm}

\renewcommand{\thefootnote}{\fnsymbol{footnote}}

\vskip 1truecm
\noindent
{\large {\sf \textbf{Jakob Palmkvist}\footnote{\it Also affiliated to: Department of Fundamental Physics,
  Chalmers University of Technology,\\ SE-412 96 G\"oteborg, Sweden}}}\\
\vskip 1truecm
        {\it 
        {Physique Th\'eorique et Math\'ematique\\
  Universit\'e Libre de Bruxelles \& International Solvay Institutes\\
  Boulevard du Triomphe, Campus Plaine, ULB-CP 231,\\BE-1050 Bruxelles, Belgium}\\[3mm]}
        {\tt jakob.palmkvist@ulb.ac.be} \\
\end{center}

\vskip 1cm

\centerline{\sf \textbf{
Abstract}}
\vskip .2cm

Gauge deformations of maximal supergravity in $D=11-n$ dimensions
generically give rise to a tensor hierarchy of $p$-form fields that transform in specific representations of the global symmetry group $E_n$. We derive the formulas defining the hierarchy from a Borcherds superalgebra corresponding to $E_n$. 
This explains why the $E_n$ representations in the tensor hierarchies also appear in the level decomposition of the 
Borcherds superalgebra. 
We show that the indefinite Kac-Moody algebra $E_{11}$ can be used equivalently to determine these representations, up to $p=D$, and for arbitrarily large $p$ if 
$E_{11}$ is replaced by $E_{r}$ with sufficiently large rank $r$.

\newpage

\pagestyle{plain}

\section{Introduction}

Eleven-dimensional supergravity is the low energy limit of M-theory, and leads by toroidal reductions to maximal supergravity in lower dimensions. General gauge deformations of the lower-dimensional theories have been systematically studied
in recent years as a way of exploring the M-theory degrees of freedom beyond supergravity \cite{deWit:2008ta,deWit:2009zv}.  
These studies have exhibited features that also appear in other approaches to M-theory, developed during the last decade,
where possible symmetries based on Kac-Moody or Borcherds algebras have been investigated 
\cite{West:2001as,HenryLabordere:2002dk,Damour:2002cu}. The present work is an attempt to relate the different approaches to each other via the features that they share.

We will in this paper consider maximal supergravity in $D$ spacetime dimensions, where $3 \leq D \leq 7$. This theory can be obtained by reduction of eleven-dimensional supergravity on an $n$-torus, where $n=11-D$, and 
it has a global symmetry group $\rm{G}$ with a corresponding Lie algebra $\g=E_n$ \cite{Cremmer:1978km,Cremmer:1979up}.
It contains a spectrum of dynamical $p$-form fields, which are antisymmetric tensors of rank $p=1,2,\ldots,D-2$ and transform in representations 
${\bf r}_{p}$ of $\g$.

One way to algebraically derive the representations ${\bf r}_p$ from only $\g$ and ${\bf r}_1$ is to consider all possible gauge deformations of the theory encoded by a so called embedding tensor. As shown in 
\cite{deWit:2004nw,deWit:2005hv,deWit:2008ta,deWit:2008gc,deWit:2009zv,Bergshoeff:2009ph}
this leads to a tensor hierarchy of $p$-forms which 
contains the dynamical $p$-form spectum. 
Another way is to embed $\g$ into an infinite-dimensional Lie (super)algebra, either a Borcherds algebra (which depends on 
$\mathfrak{g}$) or the indefinite Kac-Moody algebra $E_{11}$. In the level decomposition of the Borcherds algebra with respect to 
$\mathfrak{g}$, the representation content on level $p$ coincides with ${\bf r}_p$, up to level $D-2$
\cite{HenryLabordere:2002dk,HenryLabordere:2003rd,Henneaux:2010ys}. 
The same is true for $E_{11}$ if the level 
decomposition is done with respect to $\mathfrak{g} \oplus \sl_D$, and restricted to tensors that are antisymmetric under 
$\sl_D$ \cite{Riccioni:2007au,Bergshoeff:2007qi,Bergshoeff:2007vb,Riccioni:2007ni,Riccioni:2009hi,Riccioni:2009xr}.

It is remarkable that the same representations show up in both the tensor hierarchy and the level decomposition, although the approaches are seemingly unrelated.
Moreover, both the tensor hierarchy and the level decomposition can be continued to $p \geq D-1$ and in this way predict which non-dynamical 
$(D-1)$- and $D$-forms that are possible to add to the theory. Also these predictions are the same 
in the two approaches, apart from two exceptions in $D=3$.

In this paper we will explain why the tensor hierarchy and the level decomposition give 
the same result up to $D$-forms for $4 \leq D \leq 7$. The paper is organized as follows.
In section 2 we review the tensor hierarchy and in section 3 the level decompositions. Section 3 is more 
mathematical and divided into two subsections,
devoted to the Borcherds algebras and $E_{11}$, respectively. 
Our main result is presented in section 3.1, where we 
show that the tensors defining the hierarchy can be interpreted as
elements in the Borcherds algebra.
In section 3.2 we show that the Borcherds algebras and 
$E_{11}$ lead to the same $p$-form representations in the level decompositions up to $p=D$ (which has
been explained differently in 
\cite{Henneaux:2010ys}), and for arbitrarily large $p$ if 
$E_{11}$ is replaced by $E_{r}$ with sufficiently large rank $r$.
We conclude the paper in section 4.

\section{The tensor hierarchy}

In this section we will briefly review how the tensor hierarchy arises in the embedding tensor formalism of gauged supergravity. We will follow \cite{deWit:2008ta} and refer to this paper (and the references therein) for more information.

We start with the vector field in maximal supergravity in $D$ dimensions,
which transforms in a representation ${\bf r}_1$ of the global symmetry group ${\rm G}$, or of the corresponding Lie algebra $\g$.
We write the vector field as $A_\mu{}^\cM$, where the indices $\cM$ are associated to 
${\bf r}_1$ and $\mu=1,2,\ldots,D$ are the spacetime indices.
In gauged supergravity, a subgroup $\rm{G}_0$ of the global symmetry group $\rm{G}$ is promoted to a local symmetry group, 
with the vector field
as the gauge field. Accordingly, we can write the generators of the gauge group $\rm{G}_0$ as $X_\cM$,
with an ${\bf r}_1$ index downstairs.
However, the generators $X_\cM$ do not have to be independent, so the dimension of the gauge group $\rm{G}_0$ can be smaller than the dimension of ${\bf r}_1$.

We let $\alpha$
be the adjoint indices of $\rm{G}$ and we let $t_\alpha$ be its generators. Since the gauge group 
$\rm{G}_0$ is a subgroup of $\rm{G}$, the generators $X_\cM$ must be linear combinations of $t_\alpha$ and can be written
\begin{align}
X_\cM = \Theta_\cM{}^\alpha t_\alpha.
\end{align}
The coefficients of the linear combinations form a tensor $\Theta_\cM{}^\alpha$ which is called the {\it embedding tensor} since it describes how $\rm{G}_0$ is embedded into $\rm{G}$.

It follows from the index structure of the embedding tensor that
it transforms in the tensor product of $\bar{\bf r}_1$ and the adjoint
of $\g$. This tensor product decomposes into a direct sum of 
irreducible representations, and supersymmetry restricts the embedding tensor to only one or two of them. This restriction is known as the {\it supersymmetry constraint} or {\it representation constraint}. The requirement that $\rm G_0$ close within 
$\rm G$ leads to a second constraint on the embedding tensor, which is known as the {\it closure constraint} or {\it quadratic constraint} and can be written
\begin{align}
[X_\cM,X_\cN]=- (X_\cM)_\cN{}^\cP X_\cP.
\end{align}
Thus $(X_\cM)_\cN{}^\cP$ serve as structure constants for the gauge group, but because of the possible linear dependence in the set of generators, $(X_\cM)_\cN{}^\cP$ is in general not antisymmetric. Only when we contract $(X_\cM)_\cN{}^\cP$ with another 
$X_\cP$ the symmetric part vanishes.

When we gauge the theory we replace the partial derivatives with covariant ones,
\begin{align}
\partial_\mu \to D_\mu = \partial_\mu-gA_\mu{}^\cM X_\cM,
\end{align}
where $g$ is a coupling constant.
Then the field strength of $A_\mu{}^\cM$ becomes
\begin{align} \label{field-strength}
F_{\mu\nu}{}^\cP = 2\,\partial_{[\mu} A_{\nu]}{}^\cP + g (X_\cM)_\cN{}^\cP A_{[\mu}{}^\cM A_{\nu]}{}^\cN
\end{align}
but as shown in \cite{deWit:2008ta} this expression is not fully covariant. The recipe presented there 
(following \cite{deWit:2004nw,deWit:2005hv})
for regaining
covariance is to 
\begin{itemize}
\item[{(i)}] add a term to the gauge transformation of $A_\mu{}^\cM$ with a parameter $\Lambda_\mu{}^{\cM\cN}$:
\begin{align} \label{parameter}
\de A_\mu{}^\cM \rightarrow \de A_\mu{}^\cM +2g(X_{(\cM})_{\cN)}{}^\cP \Lambda_\mu{}^{\cM\cN},
\end{align}
\item[{(ii)}] add a term to the field strength of $A_\mu{}^\cM$ involving an new field $A_{\mu\nu}{}^{\cM\cN}$:
\begin{align} \label{newfield}
F_{\mu\nu}{}^\cP \rightarrow F_{\mu\nu}{}^\cP -2 g(X_{(\cM})_{\cN)}{}^\cP A_{\mu\nu}{}^{\cM\cN},
\end{align}
\item[{(iii)}] define the appropriate gauge transformation of the new field $A_{\mu\nu}{}^{\cM\cN}$.
\end{itemize}

\noindent
The new field $A_{\mu\nu}{}^{\cM\cN}$ carries two ${\bf r}_1$ indices and thus transforms under $\mathfrak{g}$ in the tensor product 
${\bf r}_1 \times {\bf r}_1$. It always occurs contracted with $(X_{(\cM})_{\cN)}{}^\cP$
and therefore only the symmetric part $({\bf r}_1 \times {\bf r}_1)_+$ of this tensor product enters.
Furthermore, the supersymmetry constraint on the embedding tensor restricts $A_{\mu\nu}{}^{\cM\cN}$ to only one of the irreducible representations within the symmetric tensor product $({\bf r}_1 \times {\bf r}_1)_+$.
This irreducible representation of $\g$ is what we call 
${\bf r}_2$.

By introducing a two-form field $A_{\mu\nu}{}^{\cM\cN}$ we solve the problem with the field strength of $A_\mu{}^\cM$, but on the other hand it leads to the same problem
for the field strength of $A_{\mu\nu}{}^{\cM\cN}$ --- it is not fully covariant. We can solve the problem in the same way as before, by
introducing yet another field, which will now be a three-form $A_{\mu\nu\rho}{}^{\cM\cN\cP}$ with three ${\bf r}_1$ indices, transforming in a representation ${\bf r}_3 \subset ({\bf r}_1)^3$ of $\mathfrak{g}$. 

The procedure that we have described can be continued until we reach the spacetime dimension in the number of antisymmetric indices. This gives a theory that is automatically consistent and gauge invariant. In the end we can set the coupling constant 
$g$ to zero but still keep the fields and parameters that we have added, and thus obtain an alternative formulation of the ungauged theory.

Each time we introduce a new $(p+1)$-form field
$A_{\mu_1\cdots\mu_{p+1}}{}^{\cM_1\cdots\cM_{p+1}}$ and a parameter $\Lambda_{\mu_1\cdots\mu_p}{}^{\cN_1\cN_2\cdots\cN_{p+1}}$  
we add a term
\begin{align}  
-gY^{\cM_1\cM_2\cdots\cM_p}{}_{\cN_1\cN_2\cdots\cN_{p+1}}\Lambda_{\mu_1\cdots\mu_p}{}^{\cN_1\cN_2\cdots\cN_{p+1}}
\end{align}
to the gauge transformation of the previous $p$-form, and a term 
\begin{align}
gY^{\cM_1\cM_2\cdots\cM_p}{}_{\cN_1\cN_2\cdots\cN_{p+1}}A_{\mu_1\cdots\mu_{p+1}}{}^{\cN_1\cN_2\cdots\cN_{p+1}}
\end{align}
to the field strength. 
The intertwiners $Y^{\cM_1\cM_2\cdots\cM_p}{}_{\cN_1\cN_2\cdots\cN_{p+1}}$ are defined recursively by 
the formula
\begin{align}
Y^{\cM_1\cM_2\cdots\cM_p}{}_{\cN_1\cN_2\cdots\cN_{p+1}}
&=-\de_{\cN_1}{}^{\la\cM_1}Y^{\cM_2\cdots\cM_p\ra}{}_{\cN_2\cdots\cN_{p+1}}\nn\\
&\quad\,-(X_{\cN_1})_{\cN_2\cdots\cN_{p+1}}{}^{\la\cM_1\cM_2\cdots\cM_p\ra} \label{y-relation}
\end{align}
where the angle brackets denote projection on ${\bf r}_p$. The lower indices of
the tensor
$Y^{\cM_1\cM_2\cdots\cM_p}{}_{\cN_1\cN_2\cdots\cN_{p+1}}$ then define ${\bf r}_{p+1} \subset {\bf r}_p \times {\bf r}_1$ so that, by definition,
\begin{align}
Y^{\cM_1\cM_2\cdots\cM_p}{}_{\cN_1\cN_2\cdots\cN_{p+1}}=Y^{\cM_1\cM_2\cdots\cM_p}{}_{\la\cN_1\cN_2\cdots\cN_{p+1}\ra}.
\end{align}
(Obviously we also have
$Y^{\cM_1\cM_2\cdots\cM_p}{}_{\cN_1\cN_2\cdots\cN_{p+1}}=Y^{\la \cM_1\cM_2\cdots\cM_p \ra}{}_{\cN_1\cN_2\cdots\cN_{p+1}}$.)
The recursion formula (\ref{y-relation}) is valid for $p\geq 2$. For $p=1$ we have 
\begin{align}
Y^\cP{}_{\cM\cN} = - (X_\cM)_\cN{}^\cP - (X_\cN)_\cM{}^\cP \label{lilla1}
\end{align} 
as we have already seen in (\ref{parameter}) and (\ref{newfield}).
(In \cite{deWit:2008ta} the symmetric part of $(X_\cM)_\cN{}^\cP$ is denoted by $Z^\cP{}_{\cM\cN}$. Thus our $Y^\cP{}_{\cM\cN}$ is the same as $-2Z^\cP{}_{\cM\cN}$ in \cite{deWit:2008ta}, which means that the two- and three-form and the corresponding parameters are also normalized differently compared to \cite{deWit:2008ta}.)
The second term in (\ref{y-relation}) is the component of $X_{\cN_1}$ in the tensor product $({\bf r}_1)^p$ projected on 
${\bf r}_p$. By the definition of a tensor product we have
\begin{align}
(X_{\cM})_{\cN_1\cN_2\cdots\cN_p}{}^{\cP_1\cP_2\cdots\cP_p}
&=(X_\cM)_{\cN_1}{}^{\cP_1}\de_{\cN_2}{}^{\cP_2} \cdots \de_{\cN_p}{}^{\cP_p}\nn\\
&\quad\,+\de_{\cN_1}{}^{\cP_1} (X_\cM)_{\cN_2}{}^{\cP_2} \de_{\cN_3}{}^{\cP_3} \cdots \de_{\cN_p}{}^{\cP_p}\nn\\
&\quad\,\qquad\cdots\nn\\
&\quad\,+\de_{\cN_1}{}^{\cP_1} \cdots \de_{\cN_{p-1}}{}^{\cP_{p-1}}  (X_\cM)_{\cN_p}{}^{\cP_p}.\label{x-relation1}
\end{align}

The formula (\ref{y-relation}) defines a sequence of 
representations ${\bf r}_p$ for all positive integers $p$ --- also for $p>D$ since no spacetime indices enter.
The only input is $\g$ itself, ${\bf r}_1$ and the representation constraint (which is needed to determine ${\bf r}_2$).
Below we list $\mathfrak{g}$ and ${\bf r}_p$ for $3 \leq D \leq 7$ and $1 \leq p \leq 7$ \cite{deWit:2008ta,deWit:2008gc,deWit:2009zv}.
\setlength{\arraycolsep}{5.5pt}
\\
{\renewcommand{\arraystretch}{1.5}
\begin{align*}
\begin{array}{|c|c|c|c|c|c|c|}
\hline
D&7&6&5&4&3\\
\hline
\g&\sl(5,\mathbb{R})&\so(5,5)&E_{6}&\,E_{7}\,&E_{8}\\
\hline
{\bf r}_1 & \overline{\bf 10} &{\bf 16}_c&\overline{\bf 27}&{\bf 56}&{\bf 248}\\
{\bf r}_2 & {\bf 5} &{\bf 10}&{\bf 27}&{\bf 133}&{\bf 3875}\\
{\bf r}_3 & \overline{\bf 5} &{\bf 16}_s&{\bf 78}&{\bf 912}&{\bf 3875}+{\bf 147250}\\
{\bf r}_4 & {\bf 10} &{\bf 45}&\overline{\bf 351}&{\bf 133}+{\bf 8645}&\\
{\bf r}_5 & {\bf 24} &{\bf 144}_s&{\bf 27}+{\bf 1728}&&\\
{\bf r}_6 & \overline{\bf 15}+{\bf 40} &{\bf 10}+ {\bf 126}_s+{\bf 320}&&&\\
{\bf r}_7 & {\bf 5}+\overline{\bf 45}+{\bf 70} &&&&\\
\hline
\end{array}
\end{align*}
}
\\
\noindent
Although no spacetime indices enter in the formula (\ref{y-relation}), the table shows that the representations know about spacetime. 
The duality between $p$-forms and $(D-2-p)$-forms is reflected by the relation 
${\bar{\bf r}_p} = {\bf r}_{D-2-p}$ --- the corresponding representations are conjugate to each other.
Furthermore, ${\bf r}_{D-2}$ is always the adjoint {\bf adj} of $\g$, and the last two representations in each column are related to the constraints of the embedding tensor: ${\bar{\bf r}_{D-1}}$ is the subrepresentation 
of ${\bf adj} \times {\bar{\bf r}_1}$ in which the embedding tensor must transform according to the supersymmetry constraint (except for an additional singlet in
the last column), and ${\bar{\bf r}_{D}}$ is the representation in which the closure constraint transforms. (In the tables in \cite{deWit:2008ta,deWit:2008gc,deWit:2009zv}, the entry corresponding to ${\bf r}_2$ for $D=3$ contains the additional singlet of the embedding tensor representation, although only ${\bf 3875}$ is included in 
${\bf r}_2$. 
There is also an issue with ${\bf r}_3$ for $D=3$ that we will discuss 
in the conclusion, section 4.)

Except for the last column, the representation ${\bf r}_p$ coincides with the content on level (minus) $p$ in the level decomposition of a certain Borcherds algebra or the Kac-Moody algebra $E_{11}$ with respect to $\g$ or $\g \oplus \sl_D$, respectively 
(in the $E_{11}$ case restricted to tensors which are antisymmetric in the $\sl_D$ indices).
Therefore a natural question is whether it is possible to derive the formula (\ref{y-relation}) from the Borcherds algebra or 
$E_{11}$. As we will see in the next section, the answer is affirmative.

\section{Level decompositions}

We will in this more mathematical section study the Lie algebra $\g$ of the global symmetry group ${\rm G}$, as a special case of a finite Kac-Moody algebra, and show how it can be extended to a Borcherds algebra or to the indefinite Kac-Moody algebra $E_{11}$. Borcherds algebras are also called {\it Borcherds-Kac-Moody (BKM) algebras} or {\it generalized Kac-Moody algebras.} They were first defined in \cite{Borcherds88} and generalized to superalgebras in \cite{Ray95}. For simplicity we will in this paper use the term Borcherds algebras also for the superalgebras, and the extension of $\g$ that we will discuss is in fact a superalgebra. We will not introduce more concepts than necessary, but refer to \cite{Kac,Ray06}
for a comprehensive account of Borcherds and Kac-Moody algebras. (As noted in \cite{Henneaux:2010ys}, footnote 8, there is an error in \cite{Ray06}, but this is not important for the cases that we consider here.)

As can be read off from the table above, $\g$ is the exceptional Lie algebra $E_8,E_7,E_6$ for $D=3,4,5$. We will now extend this notation and write $\g=E_n$, where $n=11-D$, also for $D=6$ and $D=7$. In fact $\g$ is the
split real form of the complex Lie algebra $E_{n}$, which is usually denoted by $E_{n(n)}$, but here we keep the simpler notation $E_{n}$ also for the split real form.

We recall that $E_n$, as a special case of a Kac-Moody algebra, can be constructed from its Dynkin diagram
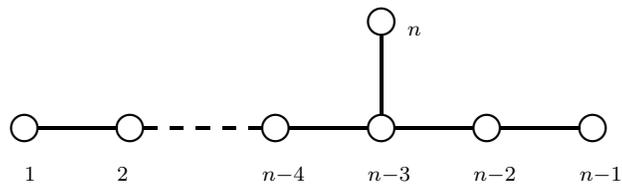
\begin{center}
\scalebox{1}{
\begin{picture}(450,60)
\put(115,-10){${\scriptstyle{1}}$}
\put(150,-10){${\scriptstyle{2}}$}
\put(205,-10){${\scriptstyle{n-4}}$}
\put(245,-10){${\scriptstyle{n-3}}$}
\put(285,-10){${\scriptstyle{n-2}}$}
\put(325,-10){${\scriptstyle{n-1}}$}
\put(260,45){${\scriptstyle{n}}$}
\thicklines
\multiput(210,10)(40,0){4}{\circle{10}}
\multiput(215,10)(40,0){3}{\line(1,0){30}}
\put(155,10){\circle{10}}
\put(115,10){\circle{10}}
\put(120,10){\line(1,0){30}}
\multiput(160,10)(10,0){5}{\line(1,0){5}}
\put(250,50){\circle{10}} \put(250,15){\line(0,1){30}}
\end{picture}}\end{center}
\vspace*{0.4cm}
by associating three {\it Chevalley generators} $e_i, f_i, h_i$ to each node ($i=1,2,\ldots,n$), satisfying the {\it Chevalley relations}
\begin{align} \label{chev-rel}
[h_i,e_j]&=A_{ij}e_j, & [h_i,f_j]&=-A_{ij}f_j, & [e_i,f_j]&=\delta_{ij}h_j, & [h_i,h_j]&=0,
\end{align}
where $A$ is the Cartan matrix corresponding to the Dynkin diagram. Any off-diagonal entry $A_{ij}$ is $-1$ if the nodes $i$ and $j$ are connected, and $0$ if they are not. The diagonal entries are all equal to $2$.

We let $\tilde{E_n}$ be the Lie algebra generated by $e_i, f_i, h_i$, and $\mathfrak{h}$ its Cartan subalgebra, spanned by 
the {\it Cartan elements} $h_i$.
Among the ideals of $\tilde{E_n}$ that intersect $\mathfrak{h}$ trivially, there is a maximal ideal,
generated by the {\it Serre relations}
\begin{align} \label{serre-rel}
(\text{ad }e_i)^{1-A_{ij}}(e_j)&=(\text{ad }f_i)^{1-A_{ij}}(f_j)=0.
\end{align}
Factoring out this ideal from $\tilde{E_n}$ we obtain the Kac-Moody algebra $E_n$.

By adding nodes to the Dynkin diagram
$E_{n}$ can be extended to a bigger algebra. 
We will in the next two subsections study two such extensions, where the extended algebra is infinite-dimensional. 
In the first case we add only one node, but we also modify the construction of the algebra and let the added node play a special role. This leads to a Borcherds algebra. In the second case we just extend the Dynkin diagram with $D=11-n$ more nodes, and accordingly we obtain the Kac-Moody algebra $E_{11}$.

\subsection{The Borcherds approach}

Following \cite{HenryLabordere:2002dk,Henneaux:2010ys} we indicate the special role of the
added node in the construction of the Borcherds algebra by painting it black, whereas the other nodes are white.
We label it by $0$, so we 
have the following
Dynkin diagram.
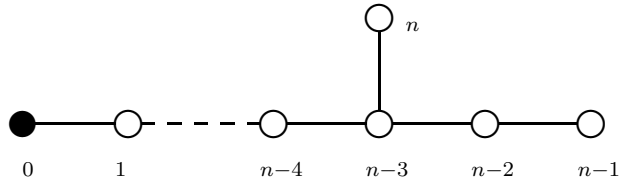
\begin{center}
\scalebox{1}{
\begin{picture}(450,60)
\put(115,-10){${\scriptstyle{0}}$}
\put(150,-10){${\scriptstyle{1}}$}
\put(205,-10){${\scriptstyle{n-4}}$}
\put(245,-10){${\scriptstyle{n-3}}$}
\put(285,-10){${\scriptstyle{n-2}}$}
\put(325,-10){${\scriptstyle{n-1}}$}
\put(260,45){${\scriptstyle{n}}$}
\thicklines
\multiput(210,10)(40,0){4}{\circle{10}}
\multiput(215,10)(40,0){3}{\line(1,0){30}}
\put(155,10){\circle{10}}
\put(115,10){\circle*{10}}
\put(120,10){\line(1,0){30}}
\multiput(160,10)(10,0){5}{\line(1,0){5}}
\put(250,50){\circle{10}} \put(250,15){\line(0,1){30}}
\end{picture}}\end{center}
\vspace*{0.4cm}
The black node plays a different role than the white ones in two respects. First, the corresponding diagonal entry in the Cartan matrix is zero, $A_{00}=0$ (instead of $A_{00}=2$). Second, the corresponding generators $e_0$ and $f_0$ are not even (bosonic) elements in an ordinary Lie algebra, but odd (fermionic) elements in a Lie {\it superalgebra}. Thus we consider the 
Lie superalgebra $\tilde{\bor}$ generated by $3(n+1)$ elements $e_I, f_I, h_I$ ($I=0,1,\ldots,\,n$) all of which are even, except for $e_0$ and $f_0$, which are odd. The Chevalley relations (\ref{chev-rel}) still hold if we generalize the ordinary antisymmetric bracket $[x,y]$ of any two elements $x$ and $y$ to a superbracket 
$\dlb x,y \drb$, which is antisymmetric if at least one of the elements is even, and symmetric if both elements are odd. In the first case we write $\dlb x,y \drb = [x,y]$ as usual, and in the second $\dlb x,y \drb = \{x,y\}$. The commutation relations among $e_0,f_0$ and $h_0$ are thus
\begin{align}
[h_0,e_0]&=[h_0,f_0]=0, & \{e_0,f_0\}&=h_0.
\end{align}
Like the Lie algebra $\tilde{E_n}$ above, which gives rise to the Kac-Moody algebra $E_n$, also the Lie superalgebra 
$\tilde{\bor}$ has a maximal ideal
that intersects the Cartan subalgebra trivially. This ideal is generated by the Serre relations 
\begin{align} \label{serre-rel}
(\text{ad }e_i)^{1-A_{iJ}}(e_J)&=(\text{ad }f_i)^{1-A_{iJ}}(f_J)=0,
\end{align}
where now $i=1,2,\ldots,n$ and $J=0,1,\ldots,n$ \cite{Ray00,Ray06}. Factoring out this ideal we obtain
a {Borcherds (super)algebra} that we here denote by $\bor_{n+1}$. 
With the notation in \cite{Henneaux:2010ys}, we have ${\bor}_{n+1}=\mathfrak{V}_{D}$.

The black node gives rise to a $\mathbb{Z}$-grading of ${\bor}_{n+1}$ which is consistent with the $\mathbb{Z}_2$-grading that 
${\bor}_{n+1}$ naturally is equipped with as a superalgebra. This means that it can be written as a direct sum of subspaces 
$({\bor}_{n+1})_{p}$ for all integers $p$, such that 
\begin{align}
\dlb ({\bor}_{n+1})_{p},({\bor}_{n+1})_{q} \drb \subseteq ({\bor}_{n+1})_{p+q}, \label{gradering}
\end{align}
where $({\bor}_{n+1})_{p}$ consists of odd elements if $p$ is odd, and of even elements if $p$ is even. 
In the grading associated to the black node, $e_0$ belongs to $({\bor}_{n+1})_{-1}$, whereas $f_0$ belongs to 
$({\bor}_{n+1})_{1}$, and all other Chevalley generators belong to $({\bor}_{n+1})_{0}$. It follows that $({\bor}_{n+1})_{0}$, as a vector space, is the direct sum of $E_{n}$ (with the Dynkin diagram obtained by removing the black node) and a one-dimensional algebra spanned by $h_0$. As a basis element of $({\bor}_{n+1})_{0}$, the Cartan element $h_0$ does not commute with $E_{n}$, but can be replaced with another Cartan element $h$ that does. This element is in the case 
$n=8$ ($D=3$) given by
\begin{align} \label{central}
c=h_0+2h_1+3h_2+4h_3+5h_4+6h_5+4h_6+2h_7+3h_8.
\end{align}
In the cases $4 \leq n \leq 7$ the Cartan element $h$ is obtained by removing the $(8-n)$ leftmost terms on the right hand side of 
(\ref{central}), and relabelling the nodes according to the Dynkin diagram above. 

According to (\ref{gradering}) 
any subspace $({\bor}_{n+1})_{p}$ closes under the adjoint action of $({\bor}_{n+1})_{0}$, and in particular of $E_n$.
Thus it constitutes a representation of $E_n$, which we call ${\bf s}_p$. Such a decomposition of (the adjoint action of) a graded Lie (super)algebra is usually called a {\it level decomposition}, where $p$ is the {\it level} of the representation ${\bf s}_p$. Determining the representations ${\bf s}_p$ explicitly in the different cases, one finds that they coincide with the 
representations ${\bf r}_p$ in the table above for $1 \leq p \leq D$, except for $D=3$ (where one in addition finds
a singlet on level 2, and an adjoint $E_8$ representation on level 3). For any level $p$, the representation ${\bf s}_{-p}$ is the conjugate of ${\bf s}_{p}$.

Starting with the fact that ${\bf r}_{1}={\bf s}_{1}$, we write the basis elements of $({\bor}_{n+1})_{-1}$ and $({\bor}_{n+1})_{1}$ as $E_\cM$ and $F^\cM$, respectively.
For $p \geq 2$ the subspace $({\bor}_{n+1})_{-p}$ is spanned by the elements
\begin{align} \label{e-uttryck}
E_{\cM_1\cdots\cM_p} \equiv \dlb E_{\cM_1},\dlb E_{\cM_2},\ldots, \dlb E_{\cM_{p-1}},E_{\cM_p} \drb \cdots \drb\drb
\end{align} 
and $({\bor}_{n+1})_{p}$ by the elements
\begin{align} \label{f-uttryck}
F^{\cM_1\cdots\cM_p} \equiv \dlb F^{\cM_1},\dlb F^{\cM_2},\ldots, \dlb F^{\cM_{p-1}},F^{\cM_p} \drb \cdots \drb\drb.
\end{align}
For any irreducible representation ${{\bf s}}$ within the tensor product $({\bf s}_1)^p$ we now want to know whether 
${{\bf s}}$ is contained in ${{\bf s}}_p$ or not. In other words, 
we want to know whether the expressions (\ref{e-uttryck})--(\ref{f-uttryck}) vanish if we project them on ${{\bf s}}$.
The lemma below is useful for
determining whether the projected expression is zero or not, 
but first we need to introduce one more concept.

In the Borcherds algebra ${\bor}_{n+1}$ we introduce a bilinear form, which we write as $\la x| y \ra$ for two elements
$x$ and $y$,
and define by
\begin{align} \label{def-innerprod}
\la h_i | h_j \ra &= A_{ij}, &\la e_i | f_j \ra &= \delta_{ij}, & \la e_i | e_j \ra=\la f_i | f_j \ra=0.
\end{align}
The definition can then be extended to the full algebra 
${\bor}_{n+1}$ in a way such that the bilinear form is invariant and supersymmetric,
\begin{align}
\la\dlb x,y \drb | z\ra &= \la x | \dlb y , z \drb \ra,&
2 \la x |y \ra &= \la x|y\ra +(-1)^{pq} \la y|x\ra,
\end{align}
where in the second equation $x \in ({\bor}_{n+1})_p$ and $y \in ({\bor}_{n+1})_q$, 
and furthermore
satisfies $\la ({\bor}_{n+1})_p | ({\bor}_{n+1})_q \ra = 0$
whenever $p+q \neq 0$. For level $\pm 1$ we have $\la E_\cM | F^\cN \ra=\de_\cM{}^\cN$.
With this bilinear form at hand we are now ready for the lemma.

\begin{lemma}
Let $x$ be an element in $({\bor}_{n+1})_{-p}$ for any $p$. Then $x=0$ if and only if $\dlb x,y \drb=0$ for all 
$y \in ({\bor}_{n+1})_{p-1}$.
\end{lemma}

\Pf If $\dlb x,y \drb=0$ for all 
$y \in ({\bor_{n+1}})_{p-1}$, then also $\la \dlb x,y \drb | z \ra=0$ for all $z$ in $({\bor_{n+1}})_{1}$, and 
by invariance of the inner product
$x$ must belong to the ideal in ${\bor}_{n+1}$ consisting of elements $u$ such that $\la u | {\bor}_{n+1} \ra=0$.
This ideal intersects the Cartan subalgebra trivially, and by the construction of ${\bor_{n+1}}$ it follows that $x=0$.
The other part of the lemma is trivial.
\qed

\noindent
The lemma says that we can as well study the expression $\dlb E_{\cN_1\cdots\cN_{p+1}}, F^{\cP_1\cdots\cP_p}\drb$ instead of 
$E_{\cN_1\cdots\cN_{p+1}}$ directly, in order to know if this is zero or not. We can then use the Jacobi identity subsequently to replace $\dlb E_{\cN_1\cdots\cN_{p+1}}, F^{\cP_1\cdots\cP_p}\drb$ by expressions that only involve lower (positive and negative) levels, until we are left with a (zero or nonzero) linear combination of the basis elements 
$E_\cM$ of $(U_{n+1})_{-1}$. The Jacobi identity for the Lie superalgebra $U_{n+1}$ can be written
\begin{align}
\dlb \dlb x,y\drb,z\drb =\dlb x, \dlb y,z \drb\drb-(-1)^{pq}\dlb y,\dlb x,z \drb\drb
\end{align}
where $x \in (U_{n+1})_p$ and $y \in (U_{n+1})_q$.
Applying it subsequently to expressions of the form $\dlb E_{\cN_1\cdots\cN_{p+1}}, F^{\cP_1\cdots\cP_p}\drb$ is of course a tedious task, but the theorem below gives a number of identities which simplify it, although they only hold under a certain condition.

For any $p \geq 2$ 
we write the projector 
corresponding to ${\bf s}_p$ 
as $(\mathbb{P}_p)_{\cM_1\cdots\cM_p}{}^{\cN_1\cdots\cN_p}$, 
with the indicies such that
\begin{align}
(\mathbb{P}_p)_{\cM_1\cdots\cM_p}{}^{\cN_1\cdots\cN_p}
&=(\mathbb{P}_p)_{\cM_1\la\cM_2\cdots\cM_p\ra}{}^{\cN_1\la\cN_2\cdots\cN_p\ra},
\end{align}
where the angle brackets denote projection on ${\bf s}_{p-1}$.
Thus we have for example
\begin{align}
E_{ \cM_1\cdots\cM_p } &= (\mathbb{P}_p)_{\cM_1\cdots\cM_p}{}^{\cN_1\cdots\cN_p} E_{\cN_1\cdots\cN_p}=
E_{\la \cM_1\cdots\cM_p \ra},\nn\\
F^{\cN_1\cdots\cN_p } &= (\mathbb{P}_p)_{\cM_1\cdots\cM_p}{}^{\cN_1\cdots\cN_p} F^{\cM_1\cdots\cM_p}=
F^{\la \cN_1\cdots\cN_p \ra}.
\end{align}

\begin{theorem}
Let $p \geq 2$ be an integer.
If there are real numbers $a_k$ such that
\begin{align}
\la E_{\cM_1 \cdots \cM_k} | F^{\cN_1 \cdots \cN_k} \ra = a_k (\mathbb{P}_k)_{\cM_1 \cdots \cM_k}{}^{\cN_1 \cdots \cN_k}
\end{align}
for $k=2,3,\ldots,p$, then the following identities hold:
\begin{align}
\dlb E_\cM, F^{\cN_1 \cdots \cN_p}\drb &= (-1)^p 
\frac{a_p}{a_{p-1}} \label{l|l-1}
\de_{\la \cM}{}^{\la \cN_1} F^{\cN_2 \cdots \cN_p\ra},
\end{align}
\begin{align}
\dlb E_{\cM_2\cdots \cM_{p}},F^{\cN_1 \cdots \cN_{p}}\drb &= 
{a_p}\,F^{\la \cN_1}\de_{\cM_2}{}^{\cN_2}\cdots\de_{\cM_p}{}^{\cN_p\ra}, \label{l|1}
\end{align}
\begin{align}
\dlb E_{\cM_1\cdots \cM_{p}},F^{\cN_1 \cdots \cN_{p}}\drb &=
{a_p}\Big(\{E_{\la \cM_1},F^{\cN_1}\}\de_{\cM_2}{}^{\cN_2}\cdots\de_{\cM_p\ra}{}^{\cN_p}\nn\\
&\qquad\quad+\de_{\la\cM_1}{}^{\cN_1} \{E_{\cM_2},F^{\cN_2}\} \de_{\cM_3}{}^{\cN_3} \cdots \de_{\cM_p\ra}{}^{\cN_p}\nn\\
&\qquad\qquad\,\cdots\nn\\
&\qquad\quad+\de_{\la\cM_1}{}^{\cN_1} \cdots \de_{\cM_{p-1}}{}^{\cN_{p-1}}  \{E_{\cM_p\ra},F^{\cN_p}\}\Big). \label{l|l}
\end{align}
\end{theorem}

\noindent
Note that if ${\bf s}_k$ is irreducible, then there must be such a number $a_k$, but not necessarily if ${\bf s}_k$ is a direct sum of irreducible representations, since the 
corresponding projectors can come with different coefficients.
\\\\
\Pf
By the assumptions we have
\begin{align}
F^{\cN_1 \cdots \cN_p}= \frac1{a_p}\la E_{\cM_1 \cdots \cM_p} |F^{\cN_1 \cdots \cN_p} \ra 
F^{\cM_1 \cdots \cM_p},
\end{align}
and by linearity we get
\begin{align}
x= \frac1{a_p}\la E_{\cN_1 \cdots \cN_p} | x \ra 
F^{\cN_1 \cdots \cN_p}
\end{align}
for any $x \in (\bor_{n+1})_{p}$. In particular
\begin{align}
\dlb E_\cM, F^{\cN_1 \cdots \cN_p}\drb &= 
\frac1{a_{p-1}}
\la E_{\cM_2 \cdots \cM_p} | \dlb E_\cM, F^{\cN_1 \cdots \cN_p}\drb \ra F^{\cM_2 \cdots \cM_p}
\nn\\
&= (-1)^p\frac1{a_{p-1}}
\la E_{\cM \cM_2 \cdots \cM_p} | F^{\cN_1 \cdots \cN_p}\ra F^{\cM_2 \cdots \cM_p}\nn\\
&= (-1)^p\frac{a_p}{a_{p-1}}
(\mathbb{P}_p)_{\cM\cM_2 \cdots \cM_p}{}^{\cN_1 \cdots \cN_{p}} F^{\cM_2 \cdots \cM_p}\nn\\
&= (-1)^p \frac{a_p}{a_{p-1}} 
\de_{\cM}{}^{\la \cN_1} F^{\cN_2 \cdots \cN_p\ra},
\end{align}
which gives (\ref{l|l-1}), and
\begin{align}
\dlb E_{\cM_2\cdots \cM_{p}},F^{\cN_1 \cdots \cN_{p}}\drb &= 
\la E_{\cM_1} | \dlb E_{\cM_2\cdots \cM_{p}},F^{\cN_1 \cdots \cN_{p}}\drb  \ra F^{\cM_1}\nn\\
&= 
\la E_{\cM_1 \cdots \cM_p} | F^{\cN_1 \cdots \cN_{p}}\ra F^{\cM_1}
\nn\\
&= {a_p}
(\mathbb{P}_p)_{\cM_1 \cdots \cM_p}{}^{\cN_1 \cdots \cN_{p}} F^{\cM_1}
\nn\\
&= 
{a_p} F^{\la \cN_1}\de_{\cM_2}{}^{\cN_2}\cdots\de_{\cM_p}{}^{\cN_p\ra},
\end{align}
which gives (\ref{l|1}).
We can now prove the identity (\ref{l|l}) by induction. It is trivially true already for $p=1$. Suppose that it holds when $p=q-1$, for some $q\geq 2$. Then using (\ref{l|l-1}) and (\ref{l|1}) we obtain
\begin{align}
\dlb E_{\cM_1\cdots \cM_{q}},F^{\cN_1 \cdots \cN_{q}}\drb 
&= (-1)^q\dlb E_{\cM_2\cdots \cM_{q}}, \dlb E_{\cM_1}, F^{\cN_1\cdots \cN_{q}}\drb\drb\nn\\
&\quad\,+\dlb E_{\cM_1} , \dlb E_{\cM_2\cdots \cM_{q}} , F^{\cN_1\cdots \cN_{q}}\drb\drb\nn\\
&= \frac{a_p}{a_{p-1}}
\de_{\cM_1}{}^{\la \cN_1} \dlb E_{\cM_2 \cdots \cM_p}, F^{\cN_2\cdots \cN_{q}\ra}\drb\nn\\
&\quad\,+\frac{a_q}{a_1} \{E_{\cM_1},F^{\la \cN_1}\} \de_{\cM_2}{}^{\cN_2}\cdots\de_{\cM_q}{}^{\cN_q\ra}\nn\\
&=\frac{a_q}{a_1}
\Big(\{E_{\cM_1},F^{\la \cN_1}\}\de_{\cM_2}{}^{\cN_2}\cdots\de_{\cM_q}{}^{\cN_q\ra}\nn\\
&\qquad\quad+\de_{\cM_1}{}^{\la\cN_1} \{E_{\cM_2},F^{\cN_2}\} \de_{\cM_3}{}^{\cN_3} \cdots \de_{\cM_q}{}^{\cN_q\ra}\nn\\
&\qquad\qquad\cdots\nn\\
&\qquad\quad+\de_{\cM_1}{}^{\la\cN_1} \cdots \de_{\cM_{q-1}}{}^{\cN_{q-1}}  \{E_{\cM_q},F^{\cN_q\ra}\}\Big),
\end{align} 
which gives (\ref{l|l}) for all $p\geq 2$ by the principle of induction.
\qed

\noindent
We have now arrived at the main result of this paper. 

\begin{cor}
For any integer $k \geq 1$, set
\begin{align}
{\tt Y}^{\cN_1\cdots\cN_{k}}{}_{\cM_1\cdots\cM_{k+1}} &= \frac1{a_{k}}[F^{\cN_1\cdots\cN_{k}},E_{\cM_1\cdots\cM_{k+1}}],\nn\\
{\tt X}_{\cM|\cN_1\cdots\cN_{k}}{}^{\cP_1\cP_2\cdots\cP_{k}} &= \frac1{a_{k}}[E_{\cM},
\dlb E_{\cN_1\cN_2\cdots\cN_{k}},F^{\cP_1\cP_2\cdots\cP_{k}} \drb].
\end{align}
Then, for $p \geq 2$ and with the conditions in the theorem, we have
\begin{align} 
{\tt X}_{\cM|\cN_1\cdots\cN_p}{}^{\cP_1\cP_2\cdots\cP_p}
&={\tt X}_{\cM|\cN_1}{}^{\la\cP_1}\de_{\cN_2}{}^{\cP_2} \cdots \de_{\cN_p}{}^{\cP_p\ra}\nn\\
&\quad\,+\de_{\cN_1}{}^{\la\cP_1} {\tt X}_{\cM|\cN_2}{}^{\cP_2} \de_{\cN_3}{}^{\cP_3} \cdots \de_{\cN_p}{}^{\cP_p\ra}\nn\\
&\qquad\cdots\nn\\
&\quad\,+\de_{\cN_1}{}^{\la\cP_1} \cdots \de_{\cN_{p-1}}{}^{\cP_{p-1}}  {\tt X}_{\cM|\cN_p}{}^{\cP_p\ra}\label{x-relation},
\end{align}
\begin{align}
{\tt Y}^{\cN_1\cdots\cN_p}{}_{\cM_1\cdots\cM_{p+1}}
&=-\de_{\cM_1}{}^{\la\cN_1}{\tt Y}^{\cN_2\cdots\cN_p\ra}{}_{\cM_2\cdots\cM_{p+1}}\nn\\
&\quad\,-{\tt X}_{\cM_1|\cM_2\cdots\cM_{p+1}}{}^{\cN_1\cN_2\cdots\cN_p},\label{y-relation2}
\end{align}
\begin{align}
{\tt Y}^{\cM}{}_{\cN\cP}=-{\tt X}_{\cN|\cP}{}^\cM-{\tt X}_{\cP|\cN}{}^\cM. \label{lilla}
\end{align}
\end{cor}
\Pf
The identity (\ref{x-relation}) follows directly from (\ref{l|l}), whereas the Jacobi identity gives (\ref{lilla}) and
\begin{align}
a_p {\tt Y}^{\cN_1\cdots\cN_p}{}_{\cM_1\cdots\cM_{p+1}}&=
[F^{\cN_1\cdots\cN_{p}},E_{\cM_1\cdots\cM_{p+1}}]\nn\\
&=-[E_{\cM_1},\dlb E_{\cM_2\cdots\cM_{p+1}},F^{\cN_1\cdots\cN_p}\drb]\nn\\
&\quad\,-(-1)^{p+1}[E_{\cM_2\cdots\cM_{p+1}},\dlb E_{\cM_1},F^{\cN_1\cdots\cN_p}\drb]\nn\\
&=-[E_{\cM_1},\dlb E_{\cM_2\cdots\cM_{p+1}},F^{\cN_1\cdots\cN_p}\drb]\nn\\
&\quad\,-\frac{a_p}{a_{p-1}}\de_{\cM_1}{}^{\la\cN_1}[F^{\cN_2\cdots\cN_p\ra},E_{\cM_2\cdots\cM_{p+1}}]\nn\\
&=-\,a_p{\tt X}_{\cM_1|\cM_2\cdots\cM_{p+1}}{}^{\cN_1\cN_2\cdots\cN_p}\nn\\
&\quad\,-a_p\de_{\cM_1}{}^{\la\cN_1}{\tt Y}^{\cN_2\cdots\cN_p\ra}{}_{\cM_2\cdots\cM_{p+1}},
\end{align}
using (\ref{l|l-1}) and (\ref{x-relation}). 
\qed

\noindent
The relations (\ref{x-relation})--(\ref{lilla}) are nothing but the definition (\ref{x-relation1}), the recursion formula 
(\ref{y-relation}) and the initial condition (\ref{lilla1}), 
with $X$ and $Y$ replaced by $\tt X$ and $\tt Y$.
We have thus derived these formulas from the Borcherds algebra $\bor_{n+1}$. 

It follows that if
${\bf r}_2={\bf s}_2$, then ${\bf r}_k \subseteq {\bf s}_k$ for $k=1,2,\ldots,p+1$ as long as the condition in the theorem is satisfied.
The condition ${\bf r}_2={\bf s}_2$ must be inserted by hand, since the general formulas (\ref{lilla1}) and 
(\ref{lilla}) are not enough to determine 
${\bf r}_2$ and ${\bf s}_2$ if we do not know what $(X_\cM)_\cN{}^\cP$ and $\tt X_{\cM|\cN}{}^\cP$ are.
We thus have to insert the definitions 
\begin{align}
(X_\cM)_\cN{}^\cP &= \Theta_\cM{}^\al (t_\al)_\cN{}^\cP, &    {\tt X}_{\cM|\cN}{}^\cP = [E_\cM,\{E_\cN,F^\cP\}]
\end{align}
in (\ref{lilla1}) and 
(\ref{lilla}) to see which representations ${\bf r}_2$ and ${\bf s}_2$ are. A priori they can be any parts of the
symmetric tensor products $({\bf r}_1 \times {\bf r}_1)_+$ and $({\bf s}_1 \times {\bf s}_1)_+$.
The correct representation is then singled out ultimately by the Serre relation 
\begin{align} \label{serrelationen}
\{ e_0, [ e_0, e_1 ]\}=0
\end{align}
on the Borcherds side, and by 
the supersymmetry constraint 
on the tensor hierarchy side
(the tensor product $\bar{\bf s}_2 \times {\bf s}_1$ must have a nonzero overlap with the representation to which the embedding tensor belongs).
Remarkably, the Serre relation (\ref{serrelationen}) and the supersymmetry constraint give the same result, so that ${\bf s}_2 = {\bf r}_2$, in all cases except for $n=8$ ($D=3$), where we have 
${\bf s}_2 = {\bf r}_2+{\bf 1}$. 

For $p\geq 3$ we can determine ${\bf s}_p$ from the formula (\ref{y-relation2}) without knowing what $\tt X_{\cM|\cN}{}^\cP$ is --- the Serre relation (\ref{serrelationen}) is automatically taken into account. However, on the tensor hierarchy side we cannot a priori exclude the possibility that the supersymmetry constraint removes some part of ${\bf s}_p$ (again by requiring that 
$\bar{\bf s}_p \times {\bf s}_{p-1}$ have a nonzero overlap with the representation to which the embedding tensor belongs).
Thus we can a priori only conclude ${\bf r}_k \subseteq {\bf s}_k$ for $k=1,2,\ldots,p+1$,
but in fact we have ${\bf r}_p = {\bf s}_p$
for $p=1,2,\ldots,D$, which shows that the Serre relation (\ref{serrelationen}) is really equivalent to the supersymmetry constraint in this sense.

For $D=7$ the representation ${\bf s}_6$ is reducible, and therefore the condition in the theorem is not automatically satisfied. It would still be satisfied if the projectors corresponding to the irreducible representations $\overline{\bf 15}$ and ${\bf 40}$ came with the same prefactor, but a computation of the inner product on level $\pm6$ in the Borcherds algebra $\bor_{n+1}$ shows that the relative prefactors are 3 and 4. Nevertheless, we have 
${\bf r}_6={\bf s}_6$, which means that the condition is sufficient but not necessary. 

We end this subsection by showing how the representations ${\bf s}_p$ can be determined in the general case, when the condition in the theorem is not satisfied. It is convenient to introduce the notation
\begin{align} \label{gen-structconst}
f_{\cN_1\cdots\cN_{p}}{}^{\cP_1\cdots\cP_{p}} &= 
\la E_{\cN_1\cdots\cN_{p}} | F^{\cP_1\cdots\cP_{p}}\ra\nn\\
&= (-1)^{p+1} 
\la \dlb E_{\cN_1\cdots\cN_{p}}, F^{\cP_2\cdots\cP_{p-1}} \drb | F^{\cP_1}\ra,
\end{align}
since, according to the lemma, the lower indices in (\ref{gen-structconst}) determine ${\bf s}_p$.
We also write 
\begin{align} \label{lilla-f}
f_\cM{}^\cN{}_\cP{}^\cQ=\la [\{E_\cM,F^\cN\},E_\cP] | F^\cQ \ra
\end{align}
and note that $(\bor_{n+1})_{-1}$ can be considered as a triple system with (\ref{lilla-f}) as structure constants for the triple product,
\begin{align} \label{borcherds-trippelprod}
(E_\cM,E_\cN,E_\cP) \mapsto [\{E_\cM,\sigma(E_\cN)\},E_\cP]=[\{E_\cM,F^\cN\},E_\cP] = f_\cM{}^\cN{}_\cP{}^\cQ E_\cQ,
\end{align}
where $\sigma$ is the superinvolution given by $\sigma(E_\cM)=F^\cM$ and $\sigma(F^\cM)=-E_\cM$.

As an aside, we mention that $(\bor_{n+1})_{-1}$ with the triple product (\ref{borcherds-trippelprod})
is a {\it generalized Jordan triple system}, like the {\it three-algebras} considered in for example
\cite{Bagger:2007jr,Bagger:2008se,Nilsson:2008kq}. However the triple system
$(\bor_{n+1})_{-1}$ is not an $\cN=6$ three-algebra since the triple product is not antisymmetric,
\begin{align}
f_\cM{}^\cN{}_\cP{}^\cQ \neq -f_\cP{}^\cN{}_\cM{}^\cQ
\end{align}
(and not an $\cN=5$ three-algebra either since the triple product does not satisfy the generalized antisymmetry condition in 
\cite{Kim:2010kq,Palmkvist:2011aw}).
Nevertheless, the construction of an associated Lie superalgebra from any $\cN=6$ three-algebra \cite{Palmkvist:2009qq,Cantarini:2010kg} can also be applied 
to the triple system $(\bor_{n+1})_{-1}$, and gives then back the full Borcherds algebra $\bor_{n+1}$.
The non-antisymmetry of the triple product is reflected by the fact that this Lie superalgebra is not 3-graded but decomposed into infinitely many subspaces in the $\mathbb{Z}$-grading.

By applying the Jacobi identity repeatedly, we now obtain
\begin{align} 
\dlb F^{\cN}, E_{\cM_1\cdots\cM_{p}}\drb&=
\dlb \{  F^{\cN}, E_{\cM_1} \}, E_{\cM_2 \cdots \cM_p} \drb\nn\\
&\quad\, -  \dlb E_{\cM_1} , \dlb F^{\cN} , E_{\cM_2 \cdots \cM_p} \drb \drb \nn\\
&=
\dlb \{  F^{\cN}, E_{\cM_1} \}, E_{\cM_2 \cdots \cM_p} \drb\nn\\
&\quad\, -  \dlb E_{\cM_1} , \dlb \{ F^{\cN} , E_{\cM_2}\} , E_{\cM_3 \cdots \cM_p} \drb \drb \nn\\
&\quad\, +  \dlb E_{\cM_1} , \dlb  E_{\cM_2} , \dlb  F^{\cN} , E_{\cM_3 \cdots \cM_p} \drb \drb \drb \nn\\
&=
\dlb \{  F^{\cN}, E_{\cM_1} \}, E_{\cM_2 \cdots \cM_p} \drb\nn\\
&\quad\, -  \dlb E_{\cM_1} , \dlb \{ F^{\cN} , E_{\cM_2}\} , E_{\cM_3 \cdots \cM_p} \drb \drb \nn\\
&\quad\, +  \dlb E_{\cM_1} , \dlb  E_{\cM_2} , \dlb  \{ F^{\cN} , E_{\cM_3}\}, E_{\cM_4 \cdots \cM_p} \drb \drb \drb \nn\\
&\qquad\cdots\nn\\
&\quad\, +(-1)^{p+1}  \dlb E_{\cM_1} , \dlb  E_{\cM_2} , \ldots , \dlb E_{\cM_{p-1}}, \{F^{\cN} ,  E_{\cM_p}\} \drb \cdots \drb \drb\nn\\
&= \sum_{i=1}^{p-1} \sum_{j=i+1}^p (-1)^{i+1} f_{\cM_i}{}^\cN{}_{\cM_j}{}^\cP
E_{\cM_1 \cdots \cM_{i-1}\cM_{i+1}\cdots \cM_{j-1}\cP\cM_{j+1}\cdots \cM_p}\nn\\
&\qquad\quad\quad+(-1)^{p} f_{\cM_p}{}^\cN{}_{\cM_{p-1}}{}^\cP
E_{\cM_1 \cdots \cM_{p-2}\cP} \label{F1Ep}
\end{align}
and then, using the invariance of the inner product,
\begin{align}
f_{\cM_1\cdots\cM_{p}}{}^{\cN_1\cdots\cN_{p}}&=
\la E_{\cM_1\cdots\cM_{p}} | F^{\cN_1\cdots\cN_{p}}\ra\nn\\
&=(-1)^{p+1} \la \dlb F^{\cN_1}, E_{\cM_1\cdots\cM_{p}}\drb | F^{\cN_2\cdots\cN_{p}}\ra\nn\\
&=\sum_{i=1}^{p-1} \sum_{j=i+1}^p (-1)^{i+p}
f_{\cM_i}{}^{\cN_1}{}_{\cM_j}{}^\cP
f_{\cM_1 \cdots \cM_{i-1}\cM_{i+1}\cdots \cM_{j-1}\cP\cM_{j+1}\cdots \cM_p}{}^{\cN_2\cdots\cN_p}\nn\\
&\qquad\qquad\qquad-f_{\cM_p}{}^{\cN_1}{}_{\cM_{p-1}}{}^\cP
f_{\cM_1 \cdots \cM_{p-2}\cP}{}^{\cN_2\cdots\cN_p}. \label{borcherdsutrakning}
\end{align}

\subsection{The $E_{11}$ approach}

We will in this subsection go from the Borcherds algebra $\bor_{n+1}$ to the Kac-Moody algebra $E_{11}$ in two steps: first replace the black node with an ordinary (white) one (thereby going from 
$U_{n+1}$ to $E_{n+1}$) and then add another $10-n$ nodes, each one connected to the previous one with a single line (thereby going from $E_{n+1}$ to $E_{11}$). 
First we give
the commutation relations for level 0 and $\pm 1$ in the Borcherds algebra ${\bor}_{11-n}$ that we discussed in the previous subsection.
They are
\begin{align}
\{E_\cM,\,F^\cN\}&=(t_\alpha)_\cM{}^\cN t^\alpha + \frac1{9-n}\de_\cM{}^\cN h, &
[t^\alpha,t^\beta]&=f^{\alpha\beta}{}_\ga t^\ga, & [t^\alpha,h]&=0,\nn
\end{align}
\begin{align}
[t^\alpha,E_\cM]&=(t^\alpha)_\cM{}^\cN E_\cN, & [h,E_\cM]&=-(10-n)E_\cM,\nn\\
[t^\alpha,F^\cN]&=-(t^\alpha)_\cM{}^\cN F^\cM, & [h,F^\cN]&=(10-n)F^\cN. \label{borcherds-comm-rel}
\end{align}
As before, $t_\alpha$ are the basis elements of $\g=E_{n}$, and $f_{\alpha\beta}{}^\ga$ are the corresponding structure constants. The adjoint $E_n$ indices are raised
with the inverse of the Killing form in $E_{n}$, which coincides with the restriction of the bilinear form in $U_{n+1}$. Thus we have $\la t_\al | t^\be \ra =\de_\al{}^\be$.

Inserting (\ref{borcherds-comm-rel}) into (\ref{lilla-f}) we get
an expression for
the structure constants of the triple system,
\begin{align} \label{borcherdstripleproduct}
f_\cM{}^\cN{}_\cP{}^\cQ = (t_\alpha)_\cM{}^\cN (t^\alpha)_\cP{}^\cQ - \frac{10-n}{9-n}\de_\cM{}^\cN \de_\cP{}^\cQ.
\end{align}

When we replace the black node with a white one we get 
$E_{n+1}$, which is not a (proper) Lie superalgebra but an ordinary Lie algebra. 
But still the added node gives rise to a $\mathbb{Z}$-grading and a level decomposition with respect to $E_n$, where we find 
${\bf r}_1$ and $\bar{\bf r}_1$ on level $-1$ and $1$, respectively. Thus we can write the basis elements of the subspaces
$(E_{n+1})_{-1}$ and $(E_{n+1})_{1}$ as $E_\cM$ and $F^\cM$, respectively. As for the Borcherds algebra we can consider 
the level $-1$ subspace as a triple system, this time
$(E_{n+1})_{-1}$
with the triple product
\begin{align} \label{en-tripleproduct}
(E_\cM,E_\cN,E_\cP) \mapsto [[E_\cM,\tau(E_\cN)],E_\cP]=[[E_\cM,F^\cN],E_\cP] = g_\cM{}^\cN{}_\cP{}^\cQ E_\cQ,
\end{align}
where $\tau$ is minus the Chevalley involution, given by $\tau(E_\cM)=F^\cM$ and $\tau(F^\cM)=E_\cM$.
The commutation relations 
for level 0 and $\pm 1$ are
\begin{align}
[E_\cM,\,F^\cN]&=(t_\alpha)_\cM{}^\cN t^\alpha + \frac1{9-n}\de_\cM{}^\cN h, &
[t^\alpha,t^\beta]&=f^{\alpha\beta}{}_\ga t^\ga, & [t^\alpha,h]&=0,\nn
\end{align}
\begin{align}
[t^\alpha,E_\cM]&=(t^\alpha)_\cM{}^\cN E_\cN, & [h,E_\cM]&=(8-n)E_\cM,\nn\\
[t^\alpha,F^\cN]&=-(t^\alpha)_\cM{}^\cN F^\cM, & [h,F^\cN]&=-(8-n)F^\cN.
\end{align}
Thus the 
only differences compared to the Borcherds case (\ref{borcherds-comm-rel}) are
the eigenvalues of $h$ acting on $E_\cM$ and $F^\cN$ 
(and of course that we now have a commutator instead of an anticommutator of $E_\cM$ and $F^\cN$).
It follows that the structure constants of the triple product (\ref{en-tripleproduct}) are
\begin{align} \label{originaltripleproduct2}
g_\cM{}^\cN{}_\cP{}^\cQ = (t_\alpha)_\cM{}^\cN (t^\alpha)_\cP{}^\cQ + \frac{8-n}{9-n}\de_\cM{}^\cN \de_\cP{}^\cQ.
\end{align}

Finally we consider $E_{11}$ with the following Dynkin diagram.
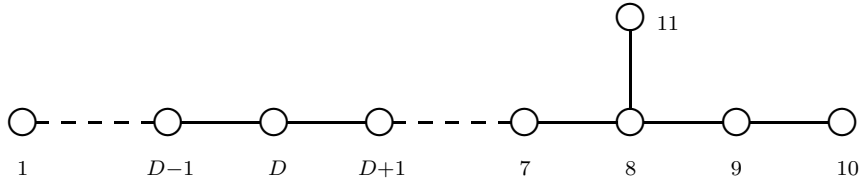
\begin{center}
\scalebox{1}{
\begin{picture}(370,60)
\put(18,-10){${\scriptstyle{1}}$}
\put(67,-10){${\scriptstyle{D-1}}$}
\put(113,-10){${\scriptstyle{D}}$}
\put(147,-10){${\scriptstyle{D+1}}$}
\put(208,-10){${\scriptstyle{7}}$}
\put(248,-10){${\scriptstyle{8}}$}
\put(288,-10){${\scriptstyle{9}}$}
\put(328,-10){${\scriptstyle{10}}$}
\put(260,45){${\scriptstyle{11}}$}
\thicklines
\multiput(210,10)(40,0){4}{\circle{10}}
\multiput(215,10)(40,0){3}{\line(1,0){30}}
\put(155,10){\circle{10}}
\put(115,10){\circle{10}}
\put(20,10){\circle{10}}
\put(75,10){\circle{10}}
\put(120,10){\line(1,0){30}}
\put(80,10){\line(1,0){30}}
\multiput(160,10)(10,0){5}{\line(1,0){5}}
\multiput(25,10)(10,0){5}{\line(1,0){5}}
\put(250,50){\circle{10}} \put(250,15){\line(0,1){30}}
\end{picture}}\end{center}
\vspace*{0.4cm}
The node that we added to the Dynkin diagram of $E_n$ in the construction of $E_{n+1}$ is now labelled $D$, and on its left hand side we have added $D-1$ more nodes, which form the Dynkin diagram of $A_{D-1}=\sl(D,\mathbb{R})=\sl_D$. This means that the node $D$ gives rise to a grading of $E_{11}$ where the subalgebra $(E_{11})_{0}$ is the direct sum of $E_n$, a one-dimensional subalgebra spanned by $h$, and $\sl_D$. It follows that any subspace $(E_{11})_{p}$ in the grading constitutes a representation of both $E_n$ and $\sl_D$.
On level $\pm 1$ we find ${\bf r}_1$ and $\bar{\bf r}_1$ as before, but now together with
the fundamental and antifundamental representations of $\sl_D$.
Thus we can write the basis elements of $(E_{11})_{-1}$ and $(E_{11})_{1}$
as $E_\cM{}^a$ and $F^\cM{}_a$, respectively, where $a=1,\ldots, D$. For $p \geq 2$, the subspace $(E_{11})_{-p}$ is then spanned by the elements
\begin{align}
E_{\cM_1\cdots\cM_p}{}^{a_1\cdots a_p} &= \dlb E_{\cM_1}{}^{a_1},\dlb E_{\cM_2}{}^{a_2},\ldots, 
\dlb E_{\cM_{p-1}}{}^{a_{p-1}},E_{\cM_p}{}^{a_p} \drb \cdots \drb\drb,
\end{align}
and $(E_{11})_{p}$ by the elements
\begin{align}
F^{\cM_1\cdots\cM_p}{}_{a_1\cdots a_p} &= \dlb F^{\cM_1}{}_{a_1},\dlb F^{\cM_2}{}_{a_2},\ldots, 
\dlb F^{\cM_{p-1}}{}_{a_{p-1}},F^{\cM_p}{}_{a_p} \drb \cdots \drb\drb.
\end{align}
If we on any level $p=1,2,\ldots,D$ 
antisymmetrize the $\sl_D$ indices, 
we find that the $E_n$ representation is the same as ${\bf s}_p$
in the corresponding level decomposition of $\bor_{n+1}$. As we saw in the preceding subsection, this representation
${\bf s}_p$ in turn coincides with ${\bf r}_p$ in the tensor hierarchy for $4 \leq D \leq 7$.
In this way the spectrum of $p$-forms that appears in gauged supergravity can be derived from $E_{11}$.
It can also be derived from $E_{10}$ in the same way if we neglect the $D$-forms, and from $E_9$ if we neglect the $(D-1)$-forms. 
Not only that, the spectrum of representations can as well be derived from $E_r$ for $r > 11$, and by choosing $r$ sufficiently large for each $p$ we get an infinite sequence of representations ${\bf t}_p$ from $E_r$, where $p$ can be any positive integer. This sequence can then be compared with the 
infinite sequences ${\bf s}_p$ and ${\bf r}_p$ coming from the Borcherds algebra and the tensor hierarchy, respectively. We end this section by showing that ${\bf s}_p={\bf t}_p$ for all $p$ (and sufficiently large $r$), and thus explaining why the Borcherds and $E_{11}$ approaches give the same result.
In \cite{Henneaux:2010ys} this has been explained in a different way, by showing that
a parabolic subalgebra of $E_{11}$ (via the tensor product with the exterior algebra, and the restriction to invariant elements) 
gives back a parabolic subalgebra of the Borcherds algebra $U_{n+1}$. The idea to consider the infinite rank extension of $E_{11}$ was also presented in \cite{Henneaux:2010ys}.

Again we consider the subspace at level $-1$ as a triple system, which in this case is $(E_{11})_{-1}$ with the
triple product
\begin{align}
(E_\cM{}^a,E_\cN{}^b,E_\cP{}^c) \mapsto
[[E_\cM{}^a,F^\cN{}_b],E_\cP{}^c] = h_\cM{}^a\,{}^\cN{}_b\,{}_\cP{}^c\,{}^\cQ{}_d\, E_\cQ.
\end{align}
As shown in \cite{Palmkvist:2007as} there is a simple formula that relates this triple product with the one (\ref{en-tripleproduct}) in $(E_{n+1})_{-1}$ above. In terms of the structure constants it reads
\begin{align} \label{simpleformula}
h_\cM{}^a\,{}^\cN{}_b\,{}_\cP{}^c\,{}^\cQ{}_d\, 
&= g_\cM{}^\cN{}_\cP{}^\cQ\, \de^a{}_b \de^c{}_d-\de_\cM{}^\cN\de_{\cP}{}^\cQ\,\de^a{}_b \de^c{}_d
+\de_\cM{}^\cN\de_{\cP}{}^\cQ\,\de^c{}_b \de^a{}_d.
\end{align}
Note that the number $11$ does not enter here --- it is only the range of the indices $a,b,\ldots=1,2,\ldots,D+r-11$ that changes when we replace 
$E_{11}$ by $E_{r}$ for some other $r \geq 12-D$.
Inserting (\ref{originaltripleproduct2}) into (\ref{simpleformula}) yields
\begin{align}
h_\cM{}^a\,{}^\cN{}_b\,{}_\cP{}^c\,{}^\cQ{}_d\,&= (t_\alpha)_\cM{}^\cN (t^\alpha)_\cP{}^\cQ \de^a{}_b \de^c{}_d \nn\\&\quad\quad+
\de_\cM{}^\cN\de_{\cP}{}^\cQ\bigg(\Big(\frac{8-n}{9-n}-1\Big)\de^a{}_b \de^c{}_d+\de^c{}_b \de^a{}_d\bigg),
\end{align}
and if we antisymmetrize in $a$ and $c$ we obtain
\begin{align} \label{observation}
h_\cM{}^{[a}\,{}^{|\cN|}{}_b\,{}_\cP{}^{c]}\,{}^\cQ{}_d\,
&=(t_\alpha)_\cM{}^\cN (t^\alpha)_\cP{}^\cQ \de^{[a}{}_b \de^{c]}{}_d +
\de_\cM{}^\cN\de_{\cP}{}^\cQ  \Big(\frac{8-n}{9-n}-2\Big)\de^{[a}{}_b \de^{c]}{}_d\nn\\
&= \de^{[a}{}_b \de^{c]}{}_d\Big( (t_\alpha)_\cM{}^\cN (t^\alpha)_\cP{}^\cQ 
-\frac{10-n}{9-n}\de_\cM{}^\cN\de_{\cP}{}^\cQ \Big)\nn\\
&=\de^{[a}{}_b \de^{c]}{}_d\,f_\cM{}^\cN{}_{\cP}{}^\cQ.
\end{align}
Thus we get back the structure constants ({\ref{borcherdstripleproduct}}) for the triple system based on the Borcherds algebra $U_{n+1}$, times $\de^{[a}{}_b \de^{c]}{}_d$.
As we will see next, this relation between the two triple systems, based on the Borcherds algebra and $E_{11}$, respectively, can be viewed as the reason why 
the two algebras give rise to the same sequence of representations.

As for the Borcherds algebra, we can define a bilinear form in $E_{11}$ (or $E_r$) by the relations (\ref{def-innerprod}). This bilinear form is still invariant, $\la [x,y]|z  \ra = \la x| [y,z] \ra$, but this time completely symmetric, since $E_{11}$ 
(or $E_r$) is an ordinary Lie algebra. By the same arguments as before, it is the lower ${\bf r}_1$ indices in
\begin{align} \label{h-innerprod}
h_{\cM_1}{}^{a_1}{}_{\cdots}{}^{\cdots}{}_{\cM_{p}}{}^{a_p}\,{}^{\cN_1}{}_{b_1}{}^{\cdots}{}_{\cdots}{}^{\cN_{p}}{}_{b_p}
= \la  E_{\cM_1\cdots\cM_p}{}^{a_1\cdots a_p}| F^{\cN_1\cdots\cN_p}{}_{b_1\cdots b_p} \ra 
\end{align}
that determine the representation ${\bf t}_p$ of $E_n$ on level $p$.
By applying the Jacobi identity repeatedly, we obtain an expression for (\ref{h-innerprod}) 
analogous to (\ref{borcherdsutrakning}).
The difference compared to (\ref{borcherdsutrakning}) is that $f$ is replaced by $h$, that each ${\bf r}_1$ index is accompanied by an $\sl_D$ index and, most important, that the prefactors $(-1)^{i+p}$ and $-1$ on the last two lines of 
(\ref{borcherdsutrakning}) are replaced by $1$ and $-1$, respectively. However, if we antisymmetrize the $\sl_D$ indices
$a_1,a_2,\ldots,a_p$ accompanying $\cM_1,\cM_2,\ldots,\cM_p$ in (\ref{borcherdsutrakning}) and use (\ref{observation}), then we can eliminate the $\sl_D$ summation index accompanying $\cP$. 
If we furthermore  
rearrange the $\sl_D$ indices in the order $a_1,a_2,\ldots,a_p$, then we pick up a factor of $(-1)^{i-1}$ on the first line and a factor of $(-1)^{p-1}$ on the second. 
After the antisymmetrization and rearrangement, the
only sign difference compared to (\ref{borcherdsutrakning}) is thus an overall factor of $(-1)^{p+1}$. 
It is then easy to show, by induction over $p$, that
\begin{align}
h_{\cM_1}{}^{[a_1}{}_{\cdots}{}^{\cdots}{}_{\cM_{p}}{}^{a_p]}\,{}^{\cN_1}{}_{b_1}{}^{\cdots}{}_{\cdots}{}^{\cN_{p}}{}_{b_p}
=(-1)^{p(p-1)/2}\de^{[a_1\cdots a_p]}{}_{b_1\cdots b_p}\,f_{\cM_1\cdots\cM_{p}}{}^{\cN_1\cdots\cN_{p}}.
\end{align}
This implies that ${\bf s}_p={\bf t}_p$ for $1 \leq p \leq D+r-11$ and all $r \geq 12-D$.

\section{Conclusion}

We have in this paper considered maximal supergravity
in $D$ spacetime dimensions, where $3 \leq D \leq 7$. We have studied three sequences of representations of $E_n$, the Lie algebra of the global symmetry group, which we have denoted by ${\bf r}_p$, ${\bf s}_p$ and ${\bf t}_p$, where $p$ can be any positive integer. The first one, ${\bf r}_p$, comes from the tensor hierarchy that arises when we gauge the supergravity theory, whereas ${\bf s}_p$ and ${\bf t}_p$ come from level decompositions of the Borcherds algebra $\bor_{n+1}$ and the Kac-Moody algebra 
$E_{11}$, respectively. It was known already that ${\bf r}_p={\bf s}_p={\bf t}_p$ for $1 \leq p \leq D-2$, and that this gives the spectrum of dynamical $p$-forms in maximal supergravity.
It was also known that ${\bf r}_p={\bf s}_p={\bf t}_p$ for $p=D-1$ and $p=D$, apart from two exceptions in the $D=3$ case. Thus the tensor hierarchy, $\bor_{n+1}$ and $E_{11}$ give the same predictions about non-dynamical $p$-forms that are possible to add to the theory. This agreement has been considered somewhat mysterious, since neither $\bor_{n+1}$ or
$E_{11}$ appears in the construction of the tensor hierarchy. On the other hand, it is in line with the ideas of gauging as a probe of M-theory degrees of freedom \cite{deWit:2008ta,deWit:2009zv}, and of Borcherds or Kac-Moody algebras as symmetries in M-theory \cite{West:2001as,HenryLabordere:2002dk,Damour:2002cu}. 

In this work we have removed much of the mystery by deriving the formulas defining the tensor hierarchy from $\bor_{n+1}$. But it is still remarkable that the supersymmetry constraint on the embedding tensor and the Serre relation (\ref{serrelationen}) on the Borcherds side restrict the symmetric tensor products $({\bf r}_1 \times {\bf r}_1)_+$ and $({\bf s}_1 \times {\bf s}_1)_+$
equally much, so that ${\bf r}_2={\bf s}_2$, in all cases except for $D=3$. It would be interesting to find an explanation also for this fact. 

Once ${\bf r}_2={\bf s}_2$, our results imply that ${\bf r}_p \subseteq {\bf s}_p$ for all $p\geq 1$ as long as the elements in 
$U_{\pm(p-1)}$ satisfy a certain condition.
Namely, their inner product must be proportional to the projector corresponding to ${\bf s}_{p-1}$. This condition is automatically satisfied if ${\bf s}_{p-1}$ is irreducible, which is the case for $2 \leq p \leq D$, except for $D=7$, where 
${\bf s}_6$ is a direct sum of two irreducible representations. We have checked that the condition is not satisfied in this case, but nevertheless, ${\bf r}_6={\bf s}_6$. Thus the condition is sufficient but not necessary for agreement. We do not know if ${\bf r}_p = {\bf s}_p$ for $4 \leq D \leq 7$ and all $p\geq 1$, also beyond the spacetime limit, but there is nothing in our results pointing in that direction.

The case $D=3$ is qualitatively different from $4 \leq D \leq 7$. In the end of section~2 we mentioned that the embedding tensor, due to the supersymmetry constraint, transforms in the representation 
${\bf r}_{D-1}$ in all cases except for $D=3$, where the embedding tensor representation contains an extra singlet missing in ${\bf r}_{D-1}={\bf r}_2$. In section 3.1 we noted that 
${\bf s}_2={\bf r}_2$ in all cases except for $D=3$, where we have ${\bf s}_2={\bf r}_2+{\bf 1}$.
The conclusion is that ${\bf s}_{D-1}$ is the representation in which the embedding tensor transforms, in {\it all} cases, even $D=3$. Since ${\bf s}_2 \neq {\bf r}_2$ for $D=3$ we cannot use the theorem and the corollary to draw any conclusions about the relation between ${\bf s}_3$ and ${\bf r}_3$. An explicit computation shows that ${\bf s}_3={\bf r}_3+{\bf 248}$, again a difference compared to $4 \leq D \leq 7$ where we have ${\bf s}_3={\bf r}_3$. It would be interesting to further investigate the role of the extra representations in ${\bf s}_2$ and ${\bf s}_3$ for $D=3$, but we leave this for future work.

The correspondence that we have derived between the components of the gauge group generators $X_\cM$
and the intertwiners $Y$ on the one hand side, and the ${\tt X}$ and ${\tt Y}$ elements 
in the subspace $(\bor_{n+1})_{-1}$ of the Borcherds algebra on the other, is not fully satisfactory. For example, the intertwiners in the tensor hierarchy satisfy an orthogonality relation 
\cite{deWit:2008ta}
\begin{align}
Y^{\cP_1\cdots\cP_p}{}_{\cN_1\cdots\cN_{p+1}}Y^{\cN_1\cdots\cN_{p+1}}{}_{\cM_1\cdots\cM_{p+2}}=0
\end{align}
but since the corresponding ${\tt Y}$ elements 
are no numbers we do not know how to multiply them with each other. 
Only expressions that are linear in ${X}$ and ${Y}$ can be translated into corresponding expressions with ${\tt X}$ and ${\tt Y}$.
It is also not clear how to interpret the $p$-form fields themselves,
the field strengths and their gauge transformations in terms of the Borcherds algebra.

Concerning the representations ${\bf t}_p$ coming from $E_{11}$ we have established that 
${\bf s}_p={\bf t}_p$ not only for $1 \leq p \leq D$, but also for arbitrarily large $p$ if we replace $E_{11}$ by $E_r$, where 
$r$ is sufficiently large. An advantage of using $E_{11}$ is that the antisymmetric $\sl_D$ indices can be naturally interpreted as the spacetime indices of the $p$-forms (if one restricts to spatial indices $E_{10}$ can be used in the same way). Another advantage is that the same algebra can be used for any $D$, we just decompose it differently, whereas the Borcherds algebra 
$U_{n+1}$ depends on $n=11-D$. On the other hand $E_{11}$ grows much faster with the levels than $U_{n+1}$, and only a tiny subset of all the tensors are antisymmetric in the $\sl_D$ indices. As long as only these tensors are of interest it is more economical to use the Borcherds algebra.

It would be interesting to in some sense extend our results to higher (and lower) dimensions.
In $D=8$ we do not expect anything special to happen, but in $D=9$ it has recently been found in
that a 9-form predicted 
by $E_{11}$ is not detected by the tensor hierarchy \cite{FernandezMelgarejo:2011wx}. In $D=10$ (type IIA and IIB) the embedding tensor formalism leading to a tensor hierarchy is not applicable, but the spectrum of $p$-form fields that can be introduced consistently with supersymmetry has been shown to agree with the predictions from the corresponding Borcherds algebras and $E_{11}$ \cite{Bergshoeff:2005ac,Bergshoeff:2006qw,Bergshoeff:2010mv,Greitz:2011da}.

\subsubsection*{Acknowledgments}
I would like to thank Martin Cederwall, Marc Henneaux, Hermann Nicolai, Bengt E.W. Nilsson, Teake Nutma, Daniel Persson, Henning Samtleben, and especially Axel Kleinschmidt for discussions and correspondence. 
The work is supported by IISN -- Belgium (conventions 4.4511.06 and 4.4514.08), by the Belgian Federal Science Policy Office 
through the Interuniversity Attraction Pole P6/11.

\bibliographystyle{utphysmod2}


\providecommand{\href}[2]{#2}
\begingroup\raggedright\endgroup


\end{document}